# Photon Tunneling Reconstitution in Black Phosphorus/*h*BN Heterostructure

Cheng–Long Zhou[1,2], Yong Zhang[1,2], Zahra Torbatian[3], Dino Novko[4,5], Mauro Antezza[6,7], and Hong–Liang Yi[1,2] *

[1]*School of Energy Science and Engineering, Harbin Institute of Technology, Harbin 150001, People's Republic of China*

[2]*Key Laboratory of Aerospace Thermophysics, Ministry of Industry and Information Technology, Harbin 150001, People's Republic of China*

[3]*School of Nano Science, Institute for Research in Fundamental Sciences (IPM), Tehran 19395–5531, Iran*

[4]*Institute of Physics, Zagreb 10000, Croatia*

[5]*Donostia International Physics Center (DIPC), Donostia – San Sebastián 20018, Spain*

[6]*Laboratoire Charles Coulomb (L2C), UMR 5221 CNRS-Université de Montpellier, F- 34095 Montpellier, France*

[7]*Institut Universitaire de France, 1 rue Descartes, F-75231 Paris, France*

**ABSTRACT**: Excitation of hybrid modes constituted by different material-supported polaritons is a common way to enhance the near-field radiative energy transport, which has fascinating promise in applications of thermal photonics. Here, we investigate near–field thermal radiation mechanisms in heterostructure composed of *h*BN film and black phosphorus single layer. The results show that this heterostructured system can give rise to a remarkable enhancement for photon tunneling, outperforming the near-field thermal radiation properties of its building blocks, as well as some other representative heterostructures. Moreover, we find that the anisotropic hybrid effect can induce a remarkable topological reconstitution of polaritons for *h*BN film and black phosphorus, forming a novel anisotropic hybrid polaritons. Notably, such hybrid modes show significant topological differences compared to *h*BN film and black phosphorus in the type-I Reststrahlen band due to the anisotropic anticrossing hybridization effect. Lastly, we systematically analyze the evolution of such hybrid polariton modes as a function of *h*BN film thickness and the corresponding influence on radiative properties of the heterostructure. This work may benefit the applications of near-field energy harvesting and radiative cooling based on hybrid polaritons in anisotropic two-dimensional material and hyperbolic film.

**KEYWORDS**: Near-field thermal radiation, Heterostructure, Anisotropy-oriented hybridization.

## I. INTRODUCTION

Compared with classical radiation [1, 2], the evanescent waves coupling between two bodies with subwavelength separation distances can assist photons to tunnel through the vacuum gap (i.e. photon tunneling), resulting in the radiative heat flux that can exceed the blackbody limit by several orders of magnitude [3, 4]. This is the so-called near-field thermal radiation (NFTR). During the last two decades, this colossal enhancement of thermal radiation has inspired a lot of research interest for its fundamental scientific relevance [5-7] and turned out to be particularly important for technological applications, such as thermal logic circuitry [8-12], near-field thermophotovoltaic [13-15], photon transformer [16], and photonic cooling [17-19]. To benefit these potential applications, continuous efforts have been devoted to explore a novel polariton modes that could intensify the photon tunneling and thermal radiation. Prominent examples include surface phonon polaritons (SPhPs) supported by polar dielectric materials such as $SiO_2$ and SiC [20-24], or surface plasmon polaritons (SPPs) that can exist at the surface of semiconductors and noble metals [25-29]. Additionally, recent advances have witnessed that hybridization of different kinds of polaritons provides an exciting paradigm to further boost the near-field thermal radiation [30-37]. In particular, researchers have found that the SPPs of graphene can couple with the multiple hyperbolic waveguide modes of the hyperbolic film to produce a fascinating dispersion behavior, which can modulate photon tunneling and provide substantial enhancement of thermal radiation on the nanoscale [38]. Inspired by this concept, many interesting ideas have been put forward in recent years, and the interested reader can consult the recent reviews of Refs. [39-45]. However, the isofrequency contour of SPPs of graphene is a closed circle, so the density of states (DOS) is finite, preventing further enhancement of photon tunneling and NFTR [46-48].

---

* Email: yihongliang@hit.edu.cn

As the technology of metamaterial fabrication develops, extensive studies have shown that compared to graphene, anisotropic 2D materials are an ideal platform for enhancing NFTR due to their higher DOS around the plasmon frequency [49-51]. A prominent example of such anisotropic 2D materials is a single layer black phosphorus (BP) or phosphorene doped with excess charge, for which radiative heat flux exceeds that of optimized graphene sheets by at least 18.5% [49]. Therefore, combining anisotropic 2D materials with the hyperbolic waveguide mode of hyperbolic films to form a novel hybrid polariton modes appears to be a promising pathway for improving NFTR [52]. Yet, unlike graphene, anisotropic 2D materials would be subject to stronger anisotropy-oriented hybridization from hyperbolic waveguide modes of hyperbolic film, significantly affecting the dispersion relation and DOS of the hybrid polaritons of heterostructure [53]. An open question is whether a novel dispersion engineering emerges under the anisotropy-oriented hybridization between hyperbolic waveguide modes and anisotropic SPPs, elevating the photon tunneling and radiative heat flux. Thus far, the influence of the strong hybridization between hyperbolic waveguide modes and anisotropic SPPs on near-field thermal radiation is not well understood.

For these reasons, in this work, we take a heterostructure composed of lightly-doped monolayer BP and hexagonal boron nitride (hBN) film as an example and reveal the role of a new anisotropic hybrid mode (hyperbolic phonon-elliptic surface plasmon polaritons) in enhancing photon tunneling and near-field thermal radiation. We theoretically predict that the anisotropy-oriented hybridization in this heterostructure not only can modulate the intensity of photon tunneling, but also enables a remarkable topological reconstitution of polaritons. Moreover, we perform the evolutionary trajectory of such hybrid polariton modes at different film thicknesses, and then investigate how this evolution modulates the radiative heat flux and photon tunneling.

## II. THEORETICAL ASPECTS

We consider the configuration of the near-field thermal radiation between two aligned heterostructures with temperatures $T_{1(2)}$=300 (310) K and vacuum gap $d$ [Fig. 1(a)]. The heterostructure consists of the 2D material with elliptic SPPs and the $h$BN film. The thickness of the $h$BN film is denoted as $t$. The elliptic surface plasmon can be obtained in many material platforms, such as from BP [49], borophene [54], germanium selenide [55], $T_d$-WTe$_2$ [56], and other 2D materials. Since BP can provide robust elliptic surface plasmon in mid-infrared region where Reststrahlen bands of $h$BN appear [57], we consider that BP is an appropriate platform to exhibit the strong hybrid effect between hyperbolic phonon polariton (waveguide) modes and elliptic surface plasmon. The optical properties of $h$BN and BP have been also addressed in Supporting Information [58].

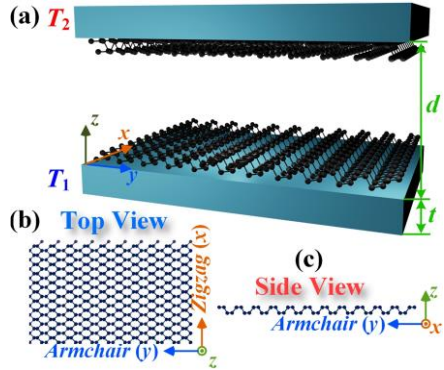

FIG. 1. (a) Schematic of near-field thermal radiation between two aligned heterostructures. Each heterostructure contains $h$BN film covered by monolayer black phosphorus. The thickness of the $h$BN film is denoted as $t$. Crystalline structure of the black phosphorus from (b) top view and (c) lateral view.

As shown in Figs. 1(b) and 1(c), the zigzag and armchair crystalline directions of BP are named as $x$- and $y$-axes, respectively. The near-field thermal radiation of such heterostructure can be characterized by the radiative heat flux (RHF). From the fluctuational electrodynamics, the radiative heat flux can be expressed as [66, 67]

$$q = \int_0^\infty g(\omega, T_1, T_2) \frac{d\omega}{2\pi} \int_0^{2\pi} \int_0^\infty \xi(\omega, k, \psi) \frac{k dk d\psi}{(2\pi)^2} \quad (1)$$

where

$$g(\omega, T_1, T_2) = \frac{\hbar\omega}{e^{\hbar\omega/k_B T_2} - 1} - \frac{\hbar\omega}{e^{\hbar\omega/k_B T_1} - 1} \quad (2)$$

here $k$ and $\Psi$ are the wavevector and the azimuthal angle in the $x$-$y$ plane, respectively. Additionally, we define the spectral radiative heat flux $q(\omega)$ as the radiative heat flux per unit of photonic energy (frequency). The $\xi(\omega,k,\Psi)$ is known as the photon transmission coefficient (PTC), implying the tunneling probability of photon. It can be written as [66]

$$\xi(\omega, k, \psi) = \begin{cases} \text{Tr}[(\mathbf{I} - \mathbf{R}^*\mathbf{R} - \mathbf{T}^*\mathbf{T})\mathbf{D}(\mathbf{I} - \mathbf{R}\mathbf{R}^* - \mathbf{T}\mathbf{T}^*)\mathbf{D}^*], & k < k_0 \\ \text{Tr}[(\mathbf{R}^* - \mathbf{R})\mathbf{D}(\mathbf{R} - \mathbf{R}^*)\mathbf{D}^*]e^{-2|k_{z0}|d}, & k > k_0 \end{cases} \quad (3)$$

where, $k_0=\omega/c$ and $k_{z0}=\sqrt{k_0^2 - k^2}$ are the wavevector and the out-of-plane wavevector in vacuum, respectively. When the in-plane wavevector $k$ is smaller than $k_0$, the electromagnetic waves excited by thermal energy are propagating mode. Otherwise, it is an evanescent wave. **I** is the identity matrix. The usual Fabry–Perot–like denominator matrix is expressed as $\mathbf{D} = (\mathbf{I} - \mathbf{R}\mathbf{R}e^{2ik_{z0}d})^{-1}$. In this expression, the reflection coefficient matrix **R** and the transmission coefficient matrix **T** have the following generic forms:

$$\mathbf{R} = \begin{bmatrix} r^{ss} & r^{sp} \\ r^{ps} & r^{pp} \end{bmatrix}; \mathbf{T} = \begin{bmatrix} t^{ss} & t^{sp} \\ t^{ps} & t^{pp} \end{bmatrix} \quad (4)$$

here, $r^{\alpha,\beta}$ and $t^{\alpha,\beta}$ with $\alpha, \beta = p, s$ are the reflection amplitude and the transmission amplitude. The reflection amplitude and transmission amplitude can be calculated by the transfer matrix methods, and the detailed derivation is given in the Supporting Information [58].

## III. RECONSTITUTION AND ENHANCEMENT OF SURFACE STATE IN HETEROSTRUCTURE

To obtain a visual evaluation for the influence of hybrid effect between $h$BN and BP on near-field thermal radiation, the RHF of three kinds of terminal structure are shown in Fig. 2(a). Three kinds of terminal structures are considered: (I) the individual BP sheet, (II) the individual $h$BN film, and (III) the heterostructure composed of BP sheet and $h$BN film. In this section, the thickness of the $h$BN film is fixed at 20 nm. As shown in Fig. 2(a), the RHF of individual black phosphorus sheet and individual $h$BN film are 60.05 kW·m$^{-2}$ and 36.67 kW·m$^{-2}$ for $d$=10 nm, respectively. Interestingly, when a monolayer BP is transferred to the surface of $h$BN film the RHF of heterostructure exceeds significantly that of other terminal structures, especially at the small vacuum gap. One can see that the heterostructure can produce $q$=164.85 kW·m$^{-2}$ for $d$=10 nm, which is more than 4.5 times larger than the RHF of the individual $h$BN film. It is noticed that, however, this enhancement diminishes significantly with the increased gap. This is because the increase in vacuum gap deteriorates the near-field coupling of the hybrid polaritons, resulting in a significant descent of evanescent contribution supported by the hybrid polaritons. As presented in Fig. 2(a), when the vacuum gap increases to 90 nm, the RHF of individual black phosphorus begins to outperform that of the heterostructure.

The application of a spectral analysis allows us to understand the mechanism for the enhanced NFTR of heterostructure. In Fig. 2(b), we show the spectral RHF of three terminal structures at $d$=10 nm. One can see that the spectral RHF of the $h$BN film mainly concentrates within the two Reststrahlen bands supported by the hyperbolic waveguide modes. Two peaks contributed by the type-I and type-II hyperbolic waveguide modes can reach 4.37 and 1.22 nW·m$^{-2}$·rad$^{-2}$·s, respectively. Fig. 2(b) shows that although the maximum value of spectral RHF is only 0.28 nW·m$^{-2}$·rad$^{-2}$·s, the broad resonant frequency range (from 0 eV/$\hbar$ to 0.3 eV/$\hbar$) enables the BP sheet to yield stronger thermal radiation. When the black phosphorus sheet is coated on the surface of $h$BN, the spectral RHF of the near-field system undergoes a significant morph, as shown in Fig. 2(b). The heterostructure combines the spectral features of $h$BN film and black phosphorus. Fig. 2(b) shows that the spectral RHF maintains relatively high values both inside and outside the Reststrahlen bands. Compared with the individual BP sheet, the introduction of $h$BN film improves the spectral characteristic of black phosphorus effectively. We take the spectral RHF at $\omega=0.959$ eV/$\hbar$ and $\omega=0.15$ eV/$\hbar$ as examples. It can be seen that the spectral RHFs at given frequencies increase respectively from 0.27 and 0.16 nW·m$^{-2}$·rad$^{-2}$·s to 1.96 and 0.92 nW·m$^{-2}$·rad$^{-2}$·s [Fig. 2(b)]. Additionally, it is worth pointing out that because of the strong hybridization effect, the spectral RHF of this heterostructure would experience a remarkable non-monotonic trend in the type-I Reststrahlen band. This also implies that a complex reconfiguration of surface state would occur in this frequency range induced by the strong hybridization effect.

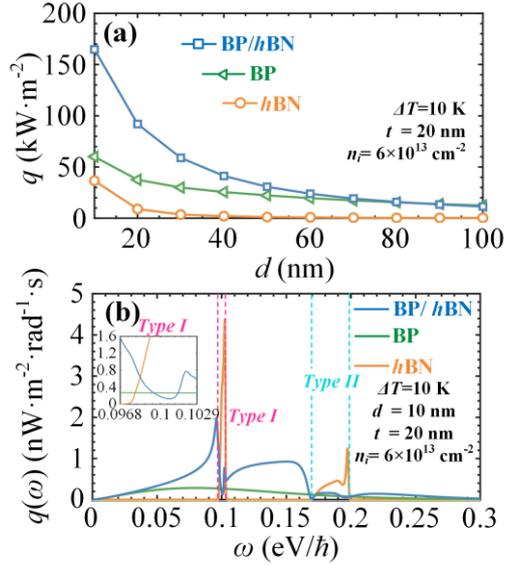

FIG. 2. (a) The RHFs for three kinds of terminal structure at different vacuum gaps. Three terminal structures are considered: (I) the individual BP sheet, (II) the individual $h$BN film, and (III) the heterostructure composed of BP sheet and $h$BN film. (b) Spectral RHFs for different terminal structures at $d=10$ nm. The thickness of the $h$BN film is fixed at 20 nm. The doping density of BP is $6 \times 10^{13}$ cm$^{-2}$.

The mechanisms behind the phenomenon of the hybridization effect can be explored by photon tunneling between the two heterostructures (i.e. photon transmission coefficient). To gain a clear evaluation of the hybridization effect, we first briefly review the surface modes of the individual $h$BN film [Fig. 3(a)] and the individual BP sheet [Fig. 3(b)]. The multiple tunneling branches resulting from the hyperbolic phonon waveguide modes of $h$BN film can be identified in the type-I and type-II Reststrahlen bands in Fig. 3(a). Its resonance branches can extend to the higher wave-vectors region around 300 $k_0$ due to the low losses within the hBN film. This is the main reason why $h$BN film can excite higher spectral RHF in the Reststrahlen bands. Additionally, the distribution of the bright branches is in good agreement with Ref. [38], which also shows that our calculations are correct. It is important to emphasize here that losses in the materials have a significant impact on near-field radiative heat transfer. Let us take the example of $h$BN film. If the losses in $h$BN are artificially ignored (that is, Im($\varepsilon_{hBN}$)=0), the material would not absorb the evanescent wave well and cannot induce an obvious near field effect of radiative heat transfer [68]. The bright branches cover a broad frequency region with a small wavevector in Fig. 3(b) for the black phosphorus sheet. This thoroughly explains the spectral feature of the black phosphorus sheet shown in Fig. 2(b).

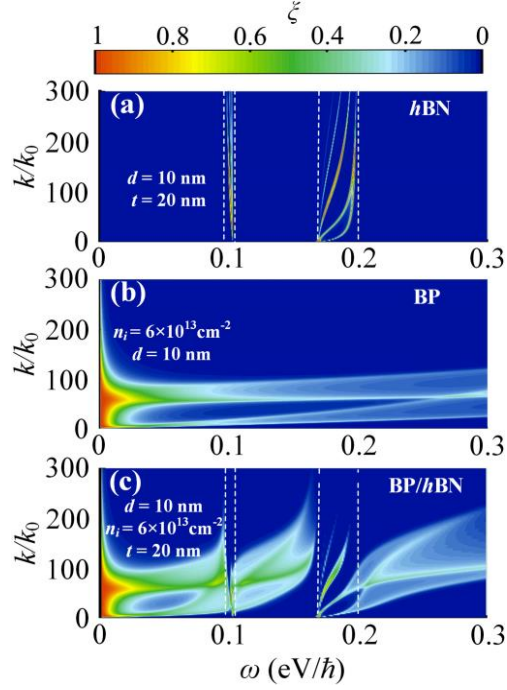

FIG. 3. The photon transmission coefficient of (a) the individual $h$BN film, (b) the individual BP sheet, and (c) the heterostructure composed of BP sheet and $h$BN film. The white dashed lines represent the Reststrahlen bands of $h$BN film. The vacuum gap, the thicknesses of $h$BN film, and the doping density of BP are fixed at 10 nm, 20 nm, and $6 \times 10^{13}$ cm$^{-2}$, respectively.

When the monolayer BP is transferred to the surface of the $h$BN film, the photon tunneling is significantly enhanced and reconfigured, as Fig. 3(c) shows. The localized fields of surface plasmons of BP outside the hyperbolic Reststrahlen bands are enhanced in the heterostructure due to the interference effect of $h$BN substrate. These intensified polaritons outside the hyperbolic Reststrahlen bands are identified as elliptical surface plasmon–phonon polaritons (ESPPPs) mode. Fig. 3(c) shows that the wavevector of ESPPPs mode can be effectively increased above 200 $k_0$ at frequencies close to the Reststrahlen bands. This strong photon tunneling may further explain the spectral enhancement outside the Reststrahlen bands shown in Fig. 2(b). The polaritons inside the two Reststrahlen bands of $h$BN are hybrid modes between elliptic SPPs and hyperbolic phonon waveguide modes, called elliptical-hyperbolic surface plasmon–phonon polaritons (EHSPPPs) mode. The hybrid effect would exhibit a significant difference in the two Reststrahlen bands. From Fig. 3(c) it is evident that the EHSPPPs mode in the type-II Reststrahlen band preserves the multiple waveguide mode features similar to the individual $h$BN film, while this feature is not evident for the EHSPPPs mode of the type-I Reststrahlen band. Furthermore, the PTCs of EHSPPPs mode in all Reststrahlen bands can still yield a robust bright branch, but its wavevector is compressed observably to a lower region. For the EHSPPPs mode of the type-I

Reststrahlen band, the wavevector of bright branches is only 100 $k_0$, as exhibited in Fig. 3(c). It is noticed that the nonlocal effects on the optical properties of materials at the large wavevectors region is an important physical phenomenon [69]. As pointed out in Ref [69-71], for the phononic materials (or 2D plasmonic monolayer material), the nonlocal correction to optical properties should be taken into account when the wavevector is greater than the inverse of the out-of-plane atomic layer spacing $a$ (or approximately $(3\sim10) \times 10^8$ m$^{-1}$). The nonlocal effects of black phosphorus and $h$BN are not considered in this work, since the wavevectors we are currently dealing with are smaller than the above wavevector range. However, when the vacuum gap is further reduced, i.e. when the wavevector range of photon tunneling is further increased, the above nonlocal effects may need to be taken into account.

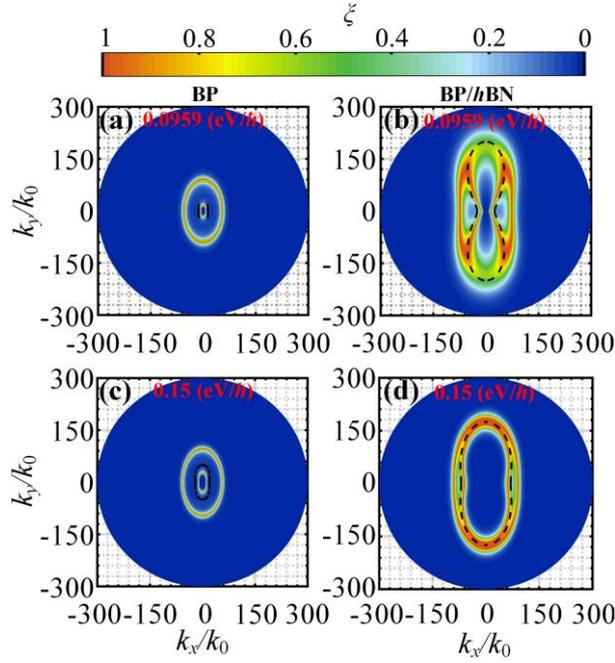

FIG. 4. In-plane PTC of (a) the black phosphorus sheet and (b) the heterostructure for the frequency of 0.0959 eV/$\hbar$. In-plane PTC of (c) the black phosphorus sheet and (d) the heterostructure for the frequency of 0.15 eV/$\hbar$. These dashed curves represent the isofrequency dispersion given in each panel.

To understand the influence of anisotropy-oriented hybridization on the photon tunneling, Figs. 4-6 shows the contours plots of in-plane PTC $\xi(\omega, k_x, k_y)$ for this heterostructure. Here, we convert $\xi(\omega, k, \Psi)$ to $\xi(\omega, k_x, k_y)$ utilizing the equations $k_x=k\cos(\Psi)$ and $k_y=k\sin(\Psi)$. Let us review the PTC features of the BP sheet in Figs. 4(a) and 4(c). At each given frequency, the PTC of BP sheets clearly shows a bright elliptical band, which matches with the results predicted by the conductivity of BP (Im[$\sigma_{xx,yy}$] > 0). Additionally, because of the coupling between the evanescent fields of the top and bottom vacuum/BP interfaces, the surface state would split into two resonant branches, i.e., the anti-symmetric and symmetric mode, as shown in Figs. 4(a) and 4(c). To confirm the dominant role of elliptical SPPs in the photon tunneling of black phosphorus, we exhibit the in-plane isofrequency dispersion at each given frequency in Figs. 4(a) and 4(c). It is shown that all dashed lines are elliptical and unambiguously located between the symmetric and anti-symmetric branches. When the monolayer black phosphorus is transferred to the surface of $h$BN film, it is first noticed that the bright branches supported by ESPPPs mode of heterostructure still show a closed topology (ellipse), as exhibited in Figs. 4(b) and 4(d). The in-plane isofrequency dispersions of ESPPPs mode of the heterostructure under all frequencies also remain elliptical, matching well with the bright branches of PTCs. This can be explained by the fact that the ESPPPs mode of the heterostructure is still dominated by BPs at these frequencies since the $h$BN film cannot provide a robust evanescent field outside the two Reststrahlen bands. Anisotropy-oriented hybridization between BP and $h$BN film cannot transform the intrinsic topological structure of BP outside the two Reststrahlen bands. However, the symmetrical and anti-symmetric resonance branches form a continuous region [Figs. 4(b) and 4(d)] since the hybrid effect in heterostructure destroys the coupling between the evanescent fields of the top and bottom BP interfaces. Additionally, compared to the elliptical SPPs of BP, ESSSPs mode of heterostructure can provide a stronger surface state at the same frequency, improving the photon tunneling of system effectively. Fig. 4 shows that the bright branches not only become more robust but also move toward a larger wavevector region. One can see that for $\omega$=0.15 eV/$\hbar$, the wavevector of ESSSPs mode can reach 240 $k_0$, which is more than two times that of the elliptical SPPs of the BP.

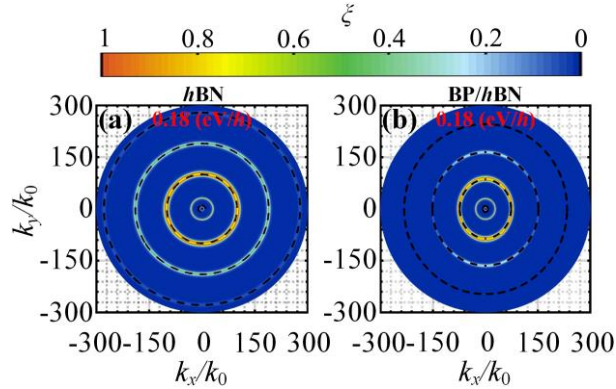

FIG. 5. In-plane PTC of (a) the $h$BN film and (b) the heterostructure at $\omega$=0.18 eV/$\hbar$. These dashed curves represent the isofrequency dispersion given in each panel.

The phenomenon of anisotropy-oriented hybridization in the two Reststrahlen bands are elaborated in the following. As a natural hyperbolic material, the $h$BN film can support hyperbolic phonon waveguide modes in two Reststrahlen bands: (i) type-I spectral regions (0.0967 - 0.1029 eV/$\hbar$) with negative out-of-plane permittivity ($\varepsilon_\perp$<0) and positive in-plane permittivity ($\varepsilon_\parallel$>0); (ii) type-II spectral regions (0.1699 - 0.1996 eV/$\hbar$) with positive out-of-plane permittivity ($\varepsilon_\perp$>0) and negative in-plane permittivity ($\varepsilon_\parallel$<0). Let us review the PTC characteristics of $h$BN film in Fig. 5(a). From the plot, it becomes apparent that multiple bright branches exist due to the Fabry-Pérot effect of hyperbolic waveguide modes in $h$BN film. Fig. 5(b) illustrates PTC of EHSPPPs mode supported by heterostructure in type-II Reststrahlen band ($\omega$=0.18 eV/$\hbar$). Here, one can see that the EHSPPPs mode in the type-II Reststrahlen band preserves the multiple resonances feature, which supports multiple bright branches. It is worth pointing out that for the type-II Reststrahlen band, the bright branches of EHSPPPs mode in the heterostructure is approximated by an ellipse [Fig. 5(b)] as opposed to the in-plane isotropy in $h$BN film [Fig. 5(a)]. Additionally, to clearly identify in-plane isotropy of EHSPPPs mode, Fig. 5(b) shows the isofrequency dispersion lines for the heterostructure. One can see that these black dashed lines are nicely located at the bright branches, which unambiguously demonstrates the anisotropic multiple resonance features of the EHSPPPs mode. This phenomenon can be understood as follows: BP has highly anisotropic constraints on the evanescent field, which allows a stronger anisotropic hybridization with the hyperbolic waveguide modes of $h$BN film and results in a higher difference in phonon waveguide propagation along the different directions. Since the greater electromagnetic loss along the zigzag crystal axis of BP slows down the wave propagation, the resonance of EHSPPPs mode at this direction is thus limited to a smaller wavevector region, as exhibited in Fig. 5(b). However, this hybrid effect does not improve the photon tunneling of such heterostructure in type-II Reststrahlen bands. Fig. 5(a) shows that the bright resonance branches of hyperbolic waveguide modes in $h$BN film can be inspired when $k$ is about 220 $k_0$. By comparison, Fig. 5(b) indicates that the anisotropy-oriented hybridization induces an early truncation of the EHSPPPs mode, and the bright resonance branches disappear at about 150 $k_0$. This early truncation also explains well why in the type-II Reststrahlen bands, the heterostructure has a lower spectral RHF compared to the $h$BN films, as indicated in Fig. 2(b).

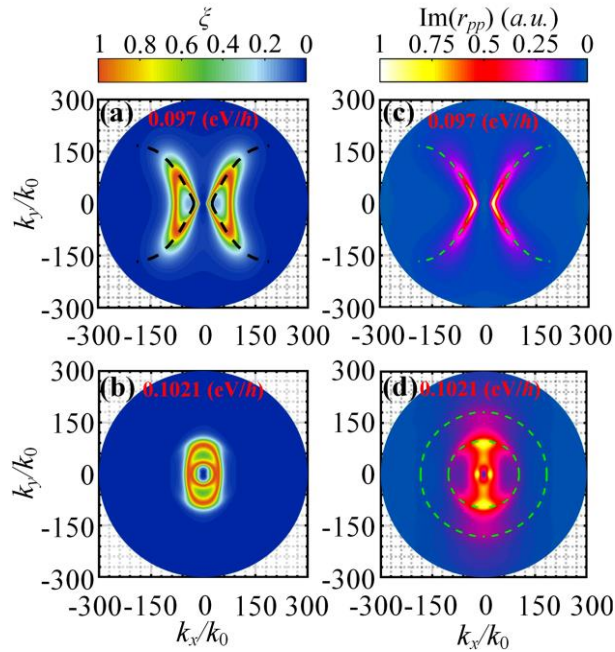

FIG. 6. In-plane PTC of the heterostructure at a frequency of (a) 0.097 eV/$\hbar$ and (b) 0.1021eV/$\hbar$. Imaginary part of the reflection coefficients Im($r_{pp}$) of the heterostructure for (c) $\omega$=0.097 eV/$\hbar$ and (d) $\omega$=0.1021 eV/$\hbar$. These dashed curves are the isofrequency dispersion given in each panel.

Interestingly, we find that the anisotropy-oriented hybridization in this heterostructure would induce a novel dispersion engineering in the type-I Reststrahlen band. Figs. 6(a) and 6(b) illustrate the contour of PTCs of the EHSPPPs mode at the frequencies of 0.097 eV/$\hbar$ and 0.1021 eV/$\hbar$, respectively. By means of the above analysis, the intrinsic topologies of both BP and $h$BN are closed. Nevertheless, as shown in Fig. 6(a), it can be clearly seen that the bright branches of the polaritons in this heterostructure perform an open topological structure (quasi-hyperbolic structure) at $\omega$=0.097 eV/$\hbar$. This is caused by the anticrossing hybridization effect between elliptical SPPs in BP and type-I hyperbolic waveguide modes in the $h$BN film. Unlike the polaritons in graphene/$h$BN heterostructure, where similar fringe patterns can be supported because of in-plane isotropic nature, the anisotropy-oriented hybridization would maximize the in-plane anisotropy, leading to a topological transition of polaritons. Similarly, in the imaginary part of the $p$-polarized reflection coefficients, there are bright hyperbolic branches, matching with the results calculated by the PTC of EHSPPPs mode. Note that since the $p$-polarized evanescent field has the main contribution to the NFTR of BP and $h$BN [38, 49-51], we only present the $p$-polarized reflection coefficients here. However, this hyperbolic property is unstable and ephemeral. As the frequency increases to 0.1021 eV/$\hbar$, the topology of the EHSPPPs model would resume to a closed structure in Figs. 6(b) and 6(d). The conductivity of black phosphorus and the permittivity of $h$BN change with the increasing of frequency, so there are many independent variables at different frequencies, which seriously interferes with qualitative analysis of why the EHSPPPs model resumes to closed topology with increasing frequency. Thus, in the subsequent section, we will adopt the thickness of $h$BN film as a single variable to elaborate in detail on the evolution of the EHSPPPs model and their topological transition processes.

## IV. EFFECT OF HBN FILM THICKNESS ON THE HYBRID POLARITONS MODE

It's worth mentioning that film thickness can strongly affect the hyperbolic phonon waveguide modes of the $h$BN film [72-74], resulting in a notable modification of anisotropy-oriented hybridization in the heterostructure. In this section, we turn now to discuss how the thickness of $h$BN films affects the thermal radiation and photon tunneling in this heterostructure. In view of the discussion above, because of the anisotropy-oriented hybridization, the EHSPPPs mode exhibits different features in contrast to the elliptical SPPs of BP and the hyperbolic waveguide modes of $h$BN, and it may contain even richer dispersion phenomena. We first investigate the thickness effect on the spectral feature of the heterostructure inside type-I hyperbolic Reststrahlen band. It is first noticed that the spectral RHF is insensitive to the variation of thickness when $t$ is large (>400 nm). Thus, the thickness of $h$BN film of Fig. 7 is restricted in the range of $t\in$[1, 500] nm. In Fig. 7, we show the thickness effect on spectral RHF of the EHSPPPs mode inside the type-I hyperbolic Reststrahlen band. There are three salient features: (i) As the thickness increases, the spectral RHF would go through three regimes, namely, quasi-elliptic regime, quasi-hyperbolic regime, and multiple resonance regime; (ii) The increase of thickness enables spectral RHF to show an increase-decrease-increase trend, in which spectral RHF would exhibit a spectral gap; (iii) The maximum value of the spectral RHF occurs when the EHSPPPs mode changes quasi-elliptic form to quasi-hyperbolic form.

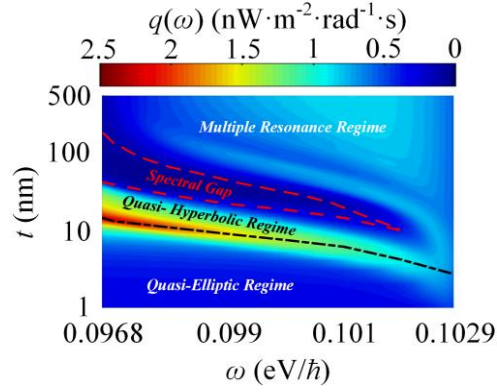

FIG. 7. Spectral RHFs for different thicknesses of $h$BN film inside the type-I hyperbolic Reststrahlen band. The vacuum gap is 10 nm. The thickness of $h$BN film is restricted in the range of $t \in [1, 500]$ nm.

Next, to better understand the mechanism of the above spectral results, we show the in-plane PTC and isofrequency dispersion with thicknesses of 1 nm, 10 nm, 12 nm, 15 nm, 25 nm, 50 nm, 100 nm, and 200 nm, respectively, in Figs. 8(a)–8(h). The frequency is fixed at 0.098 eV/$\hbar$. As one can see, the evolutionary trajectory of the EHSPPPs mode at different $h$BN film thicknesses can be clearly presented. When the thickness of $h$BN film is fixed at 1 nm, we notice that the resonance branches of the EHSPPPs mode exhibit elliptic topology. This phenomenon can be understood by the approximate dispersion relation of the independent $h$BN film, and the approximate dispersion relation can be expressed as [72]:

$$k_l = \frac{\psi}{k_0 t}\left[2\arctan\left(\frac{i}{\sqrt{\varepsilon_\perp \varepsilon_\parallel}}\right) + \pi l\right], l = 0, 1...  \quad (5)$$

where $k_l$ is the in-plane wavevector of dispersion of $h$BN film, and $\psi = i\sqrt{\varepsilon_\perp/\varepsilon_\parallel}$. According to the approximate dispersion relation, it can be derived that hyperbolic waveguide modes of $h$BN film are located at an ultrahigh wavevector regime (much larger than the wavevector of elliptic SPPs in BP), when the thickness of $h$BN film is 1 nm. Thus, for the case with $t=1$ nm, the EHSPPPs mode is governed by the evanescent field of BP, exhibiting the bright elliptic branches, as Fig. 8(a) shows. As the thickness increases to 9 nm, the dispersion of $h$BN film shrinks drastically to a smaller wavevector region, where the dispersion feature can be predicted by Eq. (5), greatly invigorating the anisotropy-oriented polaritons hybridization in the heterostructure. Moreover, the increase in thickness expands the elliptic dispersion curve governed by BP to a higher wavevector range. For the sake of subsequent discussion, we define this dispersion as a 1-order dispersion of the EHSPPPs mode. Fig. 8(b) shows that a significant dispersion variation appears for the minimum-order hyperbolic phonon waveguide mode due to the anisotropy-oriented hybridization. We note that this dispersion differs significantly from the circular guided mode of $h$BN film. It exhibits a figure-eight-like topology, and can be excited along y-axis only for the lower wavevectors. We define this dispersion as a 2-order dispersion of the EHSPPPs mode. As the thickness further increases, the above dispersion change would be more pronounced, with the 1-order and 2-order dispersion lines closer to each other along the $y$-axis. When the thickness is 12 nm, the EHSPPPs mode of heterostructure would be reconstituted topologically, exhibiting quasi-hyperbolic-like bright branches. One can see that the black dashed lines are nicely located at the bright branches, which unambiguously demonstrates the quasi-hyperbolic-like features of the EHSPPPs mode.

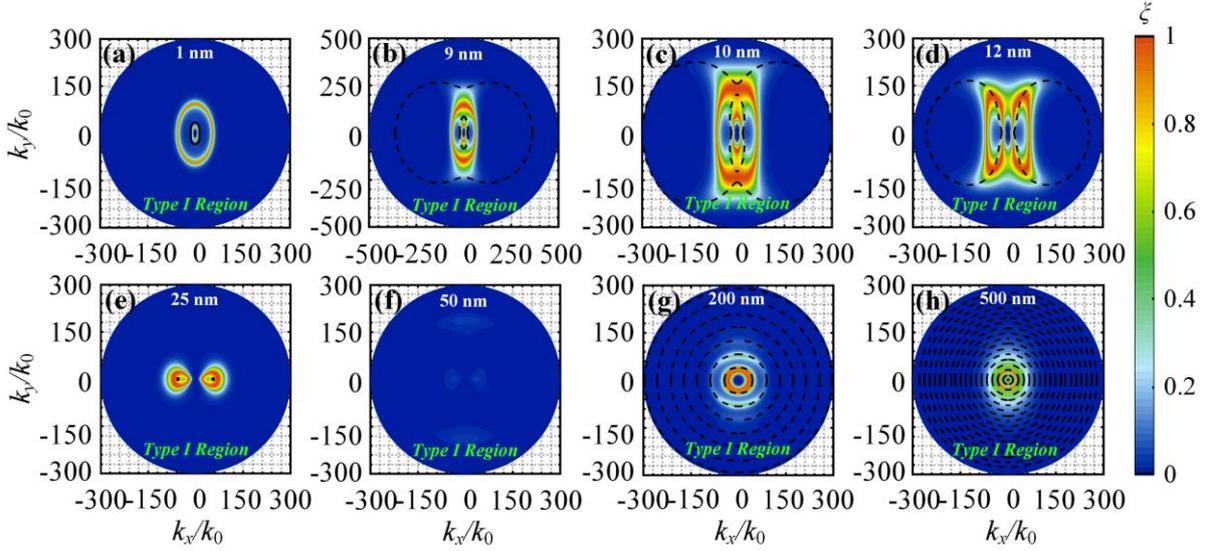

FIG. 8 In-plane PTC of the heterostructure at the frequency of 0.098 eV/$\hbar$ for (a) $t$=1 nm, (b) $t$=9 nm, (c) $t$=10 nm, (d) $t$=12 nm, (e) $t$=25 nm, (f) $t$=50 nm, (g) $t$=200 nm, and (h) $t$=500 nm. These dashed curves represent the isofrequency dispersion given in each panel.

The observed topological transition can be explained as follows: when two dispersions are infinitely close to each other, the anticrossing hybridization effect occurs since two in-plane dispersions cannot overlap, resulting in an opening topology (quasi-hyperbolic shape) of the EHSPPPs mode. This similar anticrossing effect was also found in the moiré hyperbolic metasurfaces [75-77], graphene/$h$BN heterostructure [38-40, 78] and graphene/SiC heterostructure [79, 80]. Additionally, this topological transition can extend effectively the wavevector region occupied by the EHSPPPs mode, improving the photon tunneling of the heterostructure. This also explains why the maximum spectral RHF occurs when the EHSPPPs mode changes from quasi-elliptic to quasi-hyperbolic form in Fig. 7. However, due to the mode repulsion from higher-order phonon waveguide modes of $h$BN film, the increased $t$ in Fig. 8(e) leads to the decrease in the wavevector range of quasi-hyperbolic branches. For the case with $t$=50 nm, it is difficult to observe the quasi-hyperbolic bright branches due to this mode repulsion [Fig. 8(f)], corresponding to the spectral gap of Fig. 7. When the thickness is fixed at 200 nm, the results show that there exists a strong multiple tunneling resonance in the contour of in-plane PTC. However, one can see that the isofrequency dispersion of this multiple tunneling resonance is still weakly anisotropic, unlike the multiple waveguide modes of pure $h$BN film with the in-plane isotropy. Let us take the innermost dispersion as an example in Fig. 8(g). Its wavevectors along the $y$-axis and $x$-axis are 40 $k_0$ and 45 $k_0$, respectively, exhibiting weak anisotropy. As the thickness increases to 500 nm, Fig. 8(h) shows that this multiple tunneling resonance becomes more pronounced, enhancing the photon tunneling of heterostructure. This result also explains the enhancement effect of increasing thickness on the spectral

RHF inside multiple resonances regime of Fig. 7. Meanwhile, for $t$=500 nm, the dispersion of EHSPPPs mode is also anisotropic. The wavevectors of the innermost dispersion line along the *y*-axis and *x*-axis are 15 $k_0$ and 16 $k_0$, respectively.

As discussed above, the EHSPPPs modes of the type-I and type-II Reststrahlen bands would show very different hybrid effects due to the different hyperbolic features of *h*BN in the two Reststrahlen bands. We focus here on the calculation of the spectral RHFs inside type-II Reststrahlen bands of *h*BN film at different thicknesses [see Fig. 9]. Here, we define the spectral contour inside type-II Reststrahlen bands as the quasi-elliptic regime with low spectral RHF and the multiple resonance regime with high spectral RHF. Let's take the case with $\omega$=0.18 eV/$\hbar$ as an example. When the thickness is below 8 nm, the variation of thickness does not substantially improve the spectral RHF. Within this range of $t \in$[1, 8] nm, the spectral RHF of the system is always below 0.08 nW·m$^{-2}$·rad$^{-2}$·s. It is first noticed that when the thickness is increased beyond 8 nm, there is a dramatic increase in spectral RHF, achieving 0.22 nW·m$^{-2}$·rad$^{-2}$·s at $t$=500 nm.

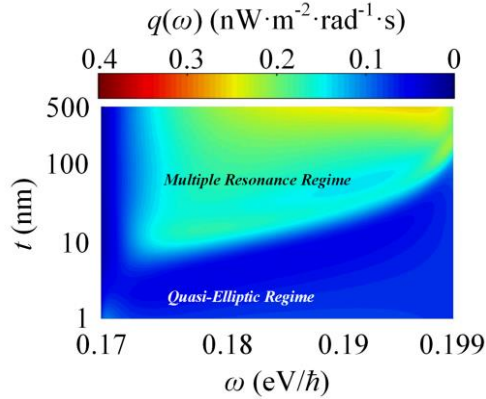

FIG. 9 Spectral RHFs for different thicknesses of *h*BN film inside the type-II hyperbolic Reststrahlen band. The vacuum gap is 10 nm. The thickness of *h*BN film is restricted in the range of $t \in$ [1, 500] nm.

Finally, we analyze the thickness effect of BP/*h*BN heterostructure inside the type-II Reststrahlen band. In Fig. 10, the in-plane PTC and isofrequency dispersion are plotted for different thicknesses at $\omega$=0.18 eV/$\hbar$. For $t$=1 nm, the hybrid effect in heterostructure is weak, and the EHSPPPs mode is governed by the evanescent field of BP, similar to the case inside type-I Reststrahlen band [Fig. 10(a)]. In Figs. 10(b)-10(d), more orders of bright branches appear with the increase in thickness and eventually merge to a continuous band when the *h*BN film is increased to 500 nm. This also explains why the thicker film is effective in improving the spectral RHF of this heterostructure in Fig. 9. Additionally, it is noticed that the resonance branches inside the type-II Reststrahlen band exhibit elliptical fringes under all thicknesses [Fig. 10], in which the optical axis of elliptical fringes is parallel to that of BP. This can be explained by the fact that the BP sheet is analogous to a converter, leading to anisotropic propagation with the elliptical fringes of the waveguide mode for *h*BN film.

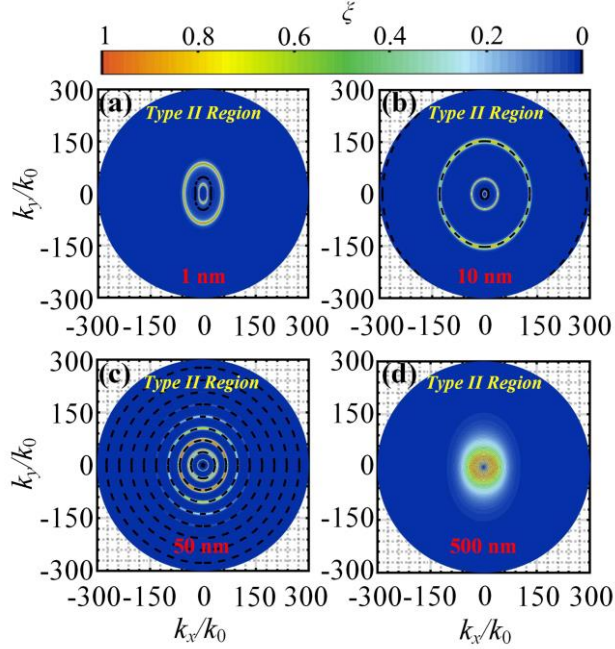

FIG. 10 In-plane PTCs of the heterostructure at the frequencies of 0.18 eV/$\hbar$ for (a) $t$=1 nm, (b) $t$=10 nm, (c) $t$=50 nm, and (d) $t$=500 nm. These dashed curves represent the isofrequency dispersion given in each panel.

## V. CONCLUSIONS

The application of hybrid modes among different polaritons has emerged as a promising way to modulate and improve near-field thermal radiation. Motivated by the current interest in the enhancement of near-field thermal radiation, we have presented in this paper a theoretical study of the anisotropic hybrid effect on the radiative heat transfer between two heterostructures composed of $h$BN film and BP. This heterostructured optical material can enhance photon tunneling and surpasses respectively individual $h$BN film and BP by a factor of 4.5 and 2.7 times in terms of achieving high near-field thermal radiation, thanks to the anisotropic hybridization effect. Moreover, we found that the elliptic surface plasmon polariton of BP can couple with the hyperbolic waveguide modes of the $h$BN film to form novel anisotropic hybrid polaritons mode, giving rise to a remarkable topological reconstitution for the polariton mode of heterostructure. Particularly, in the type-I Reststrahlen band of $h$BN, these anisotropic hybrid polaritons enable a topological transition of the surface state from elliptical (closed) to quasi-hyperbolic (open). We explain that this transition is due to the in-plane anticrossing hybridization effect. Lastly, we have systematically exhibited the evolutionary trajectory of this anisotropic hybrid polaritons mode as a function of different $h$BN film thicknesses, and then investigated how this evolution efficiently modulates the radiative heat flux and photon tunneling.

Inducing anisotropic hybrid polariton mode upon the application of the elliptic surface plasmon polariton and the hyperbolic phonon waveguide modes, as demonstrated in the present work, may also have important consequences for the enhancement and modulation of near-field radiative energy transport. The phenomena identified in this work can be effectively extended to heterostructures prepared from other 2D materials with elliptical plasmon nature (such as borophene, carbon phosphide, etc.) and other hyperbolic films (such as $Bi_2Se_3$, calcite, etc.), and open exotic avenues for the development of more efficient and powerful thermal management and photonic energy harvesting techniques.

## ACKNOWLEDGMENT

This work was supported by the National Natural Science Foundation of China (Grant No. 52076056), the Fundamental Research Funds for the Central Universities (Grant No. FRFCU5710094020), the Croatian Science Foundation (Grant no. UIP–2019–04–6869), and the European Regional Development Fund for the "Center of Excellence for Advanced Materials and Sensing Devices" (Grant No. KK.01.1.1.01.0001).

## REFRENCE